\documentclass[submission,Phys]{SciPost}
\pdfoutput=1
\usepackage{amsmath,amssymb,mathtools,xspace}
\usepackage{booktabs,multirow,graphicx,tabularx,slashed}
\usepackage{hyperref}
\usepackage{color,xcolor}
\usepackage[normalem]{ulem}
\usepackage{enumitem}
\usepackage{feynmp}
\usepackage{braket,stackrel}
\usepackage{tikz}
\usepackage{subcaption}
\usetikzlibrary{arrows}
\usetikzlibrary{shapes.geometric, arrows}

\makeatletter
\@ifundefined{pdfoutput}{}{\DeclareGraphicsRule{*}{mps}{*}{}}
\makeatother

\makeatletter
\DeclareRobustCommand*{\bfseries}{%
   \not@math@alphabet\bfseries\mathbf
   \fontseries\bfdefault\selectfont
   \boldmath
}
\makeatother

\setitemize{itemsep=2pt,topsep=2pt,parsep=0pt,partopsep=0pt,leftmargin=*}
\setenumerate{itemsep=0pt,topsep=2pt,parsep=0pt,partopsep=0pt,labelindent=3pt,leftmargin=*}

\definecolor{Gcolor}{HTML}{3b528b}
\definecolor{Dcolor}{HTML}{e41a1c}

\tikzstyle{generator} = [rectangle, rounded corners, minimum width=3cm, minimum height=1cm,text centered, draw=Gcolor]
\tikzstyle{discriminator} = [rectangle, rounded corners, minimum width=3cm, minimum height=1cm,text centered, draw=Dcolor]
\tikzstyle{io} = [circle, trapezium left angle=70, trapezium right angle=110, minimum width=1cm, minimum height=1cm, text centered, draw=black]

\tikzstyle{process} = [rectangle, minimum width=1cm, minimum height=1cm, text centered, draw=black]
\tikzstyle{decision} = [rectangle, minimum width=1cm, minimum height=1cm, text centered, draw=black]

\tikzstyle{arrow} = [thick,->,>=stealth]
\usepackage{xcolor}

\newcommand\one{\leavevmode\hbox{\small1\normalsize\kern-.33em1}}

\newcommand{\lag}{\mathcal{L}}

\newcommand{\qqquad}{\qquad \qquad}

\newcommand{\met}{\slashchar{E}_T}

\newcommand{\gev}{\text{GeV}}

\newcommand{\br}{\mathcal{B}}

\newcommand{\ifb}{\ensuremath{\text{fb}^{-1}} }

\def\slashchar#1{\setbox0=\hbox{$#1$}           
   \dimen0=\wd0                                 
   \setbox1=\hbox{/} \dimen1=\wd1               
   \ifdim\dimen0>\dimen1                        
      \rlap{\hbox to \dimen0{\hfil/\hfil}}      
      #1                                        
   \else                                        
      \rlap{\hbox to \dimen1{\hfil$#1$\hfil}}   
      /                                         
   \fi}

\begin{document}

\begin{center}{\Large \textbf{
      A Final Word on FCNC-Baryogenesis from Two Higgs Doublets
}}\end{center}

\begin{center}
  Wei-Shu Hou\textsuperscript{1},
  Tanmoy Modak\textsuperscript{2}, and
  Tilman Plehn\textsuperscript{2}
\end{center}

\begin{center}
{\bf 1} Department of Physics, National Taiwan University, Taipei, Taiwan \\
{\bf 2} Institut f\"ur Theoretische Physik, Universit\"at Heidelberg, Germany \\
modak@thphys.uni-heidelberg.de
\end{center}

\begin{center}
\today
\end{center}


\section*{Abstract}
{\bf Electroweak baryogenesis in a two-Higgs doublet model is a
  well-motivated and testable scenario for physics beyond the Standard
  Model. An attractive way of providing $CP$ violation is through
  flavor-changing Higgs couplings, where the top-charm coupling is
  hardly constrained.  This minimal scenario can be tested by searching for
  heavy charged and neutral Higgs bosons at the LHC. While the charged
  Higgs signature requires a dedicated analysis, the neutral Higgs
  signature will be covered by a general search for same-sign top
  pairs. Together, they provide a conclusive test of this kind of
  baryogenesis.}

\vspace{10pt}
\noindent\rule{\textwidth}{1pt}
\tableofcontents\thispagestyle{fancy}
\noindent\rule{\textwidth}{1pt}
\vspace{10pt}

\newpage
\section{Introduction}
\label{sec:intro}

The Higgs discovery~\cite{Aad:2012tfa,Chatrchyan:2012ufa} and
subsequent measurements of the Higgs Lagrangian at Run~2~\cite{Aad:2019mbh,Sirunyan:2018koj,Biekotter:2018rhp,Ellis:2018gqa,Almeida:2018cld,Kraml:2019sis,vanBeek:2019evb,Dawson:2020oco,Ellis:2020unq}
indicate that the Standard Model is the correct effective theory
around the electroweak scale. While there exists no experimental evidence
for physics beyond the Standard Model so far, extended Higgs sectors are
motivated by theoretical considerations, like mass generation of
up-type and down-type fermions, neutrino mass generation, electroweak
baryogenesis, or dark matter. In particular, two-Higgs doublet models
(2HDMs)~\cite{Lee:1973iz,Djouadi:2005gi,Branco:2011iw} are an integral part of
well-defined models for physics beyond the Standard Model, including
MSSM~\cite{Djouadi:2005gj}, composite Higgs
models~\cite{Hill:2002ap}, little Higgs
models~\cite{Schmaltz:2005ky,Perelstein:2005ka}, or
GUTs~\cite{Pati:1974yy,Gursey:1975ki,Achiman:1978vg,Barbieri:1981yy}.

If we use baryogenesis~\cite{Sakharov:1967dj} as a guiding principle
to new physics searches at the LHC, a 2HDM is an attractive and
minimal choice.  It can provide both, new scalar degrees of
freedom~\cite{Bochkarev:1990fx,Turok:1990zg} and $CP$-violation.  In
the general~\cite{Chen:2013qta,Chang:2017wpl} or
type-III~\cite{Hou:1991un} 2HDM, the new particles can be close in
mass to the SM-Higgs~\cite{Lopez-Val:2013yba,Haller:2018nnx}.
Sufficiently large $CP$-violation is non-trivial to achieve, in our
model we rely on the Yukawa sector.  If both doublets couple to
up-type and down-type quarks, they define two separate Yukawa
matrices. After diagonalizing the quark mass matrices we find the
real, diagonal couplings $\lambda_{ii} = \sqrt{2}m_i/v$ and the
complex, non-diagonal couplings $\rho_{ij}$.  While flavor-changing
neutral couplings are generally well-constrained, it is possible to have
electroweak baryogenesis (EWBG) driven by a single, order-one, complex
coupling $\rho_{tc}$~\cite{Fuyuto:2017ewj},
\begin{align}
  \text{Im}\, \rho_{tc} \gtrsim 0.5
  \qquad 
  \text{and}
  \qquad
  |\cos\gamma| \gtrsim 0.1 \; ,
\end{align}
where $\gamma$ is the mixing angle between the two $CP$-even Higgs
states.

It has been shown~\cite{Kohda:2017fkn,Hou:2018zmg,Hou:1997pm,Altmannshofer:2016zrn,Altmannshofer:2019ogm}
that the coupling $\rho_{tc}$ can be discovered in the LHC process
\begin{align}
  cg \to t A/H \to t \, (t \bar c) \, ,
\end{align}
where this process retains a very mild dependence on $\cos\gamma$ and
is especially useful for small values of $\cos \gamma$.  To attribute
this signal to EWBG requires information on the mixing angle
$\cos\gamma$, for instance through $b$-associated charged Higgs
production~\cite{Gori:2017tvg}
\begin{align}
  cg \to b H^+ \to b \, (W^+ h) \, .
\end{align}
Here the production process is induced by
$\rho_{tc}$~\cite{Iguro:2017ysu,Ghosh:2019exx}, while the decay
amplitude is proportional to the mixing angle $\cos \gamma$. Even in
the absence of complex phase information, such a search can test the
required particle content and parameter space for the $\rho_{tc}$-EWBG
scenario.  Finally, the exotic top decay
\begin{align}
t \to c h
\end{align}
is induced by the coupling $\rho_{tc}$ combined with non-vanishing
$\cos \gamma$~\cite{Hou:1991un} and is searched for by
CMS~\cite{Sirunyan:2017uae} and ATLAS~\cite{Aaboud:2018oqm}.

In this paper we show how the two LHC searches for charged and neutral
heavy Higgs bosons can conclusively probe the parameter region
required for $\rho_{tc}$-EWBG in the general 2HDM (g2HDM).  The paper is
organized as follows: in Sec.~\ref{sec:param} we discuss the model and
its preferred parameter space, and then compare it to the reach of the
charged Higgs channel in Sec.~\ref{sec:bwh}. Section~\ref{sec:sstop}
is dedicated to same-sign top production from neutral Higgs production
and its complementarity to the charged Higgs signature.  We summarize
our results in Sec.~\ref{sec:disc}.

\section{Model and parameter space}
\label{sec:param}

The general $CP$-conserving two Higgs doublet potential can be written
as~\cite{Davidson:2005cw,Hou:2017hiw}
\begin{align}
V(\Phi,\Phi') =\ & \mu_{11}^2|\Phi|^2 + \mu_{22}^2|\Phi'|^2
            - \left(\mu_{12}^2\Phi^\dagger\Phi' + \text{h.c.}\right)
            + \frac{\eta_1}{2}|\Phi|^4 + \frac{\eta_2}{2}|\Phi'|^4
            + \eta_3|\Phi|^2|\Phi'|^2 \notag \\
 &  + \eta_4 |\Phi^\dagger\Phi'|^2 
    + \left[\frac{\eta_5}{2}(\Phi^\dagger\Phi')^2
    + \left(\eta_6 |\Phi|^2 + \eta_7|\Phi'|^2\right) \Phi^\dagger\Phi' + \text{h.c.} \right]  .
\label{eq:pot}
\end{align}
In the Higgs basis, the VEV $v = 246$~GeV is generated by the doublet
$\Phi$, while $\Phi'$ does not develop a VEV, hence $\mu_{22}^2 > 0$.
The minimization conditions in the two field directions lead to
$\mu_{11}^2 = -\eta_1 v^2/2$ and $\mu_{12}^2 = \eta_6 v^2/2$. The
mixing angle $\gamma$ diagonalizes the $CP$-even mass matrix to define
the mass eigenstates $h$ and $H$~\cite{Davidson:2005cw,Hou:2017hiw},
\begin{align}
  c_\gamma^2 = \cos^2 \gamma = \frac{\eta_1 v^2 - m_h^2}{m_H^2-m_h^2}
  \qquad \text{and} \qquad 
  s_{2\gamma} = \sin (2 \gamma) = \frac{2\eta_6 v^2}{m_H^2-m_h^2} \; ,
\label{eq:angles}
\end{align}
where $c_\gamma \to 0$ in the alignment limit ($\eta_6 = 0$ and
$\eta_1 = m_h^2/v^2 \sim 1/4$).  To satisfy the first Sakharov
condition~\cite{Sakharov:1967dj}, a new scalar degree of freedom close
in mass with the
SM-Higgs~\cite{Basler:2018cwe,Basler:2019iuu,Basler:2020nrq} can
trigger a strong first-order phase transition. Following
Eq.\eqref{eq:angles} this is guaranteed by finite $c_\gamma$ and
perturbatively stable $\eta_i = \mathcal{O}(1)$, for instance $\eta_6
= {\cal O}(1)$ and $\eta_1 = {\cal O}(1) >
m_h^2/v^2$~\cite{Hou:2017hiw}.

Next, baryogenesis requires a complex phase in the Higgs or
Yukawa sectors~\cite{Sakharov:1967dj}. Many analyses have studied a
complex Higgs potential, which tends to be strongly constrained by EDM
measurements~\cite{Cline:1995dg,Funakubo:1995kw,Dorsch:2016nrg,Basler:2017uxn}.
We look at the alternative option of $CP$-violation arising from the
Yukawa sector~\cite{Davidson:2005cw,Fuyuto:2017ewj,Hou:2019mve}
\begin{align}
\lag \supset
& -\frac{1}{\sqrt{2}} \sum_{F = U,D,L} \overline F_{i} \Big[
  \left( -\lambda^F_{ij} s_\gamma + \rho^F_{ij} c_\gamma\right) h 
  +\left( \lambda^F_{ij} c_\gamma + \rho^F_{ij} s_\gamma\right)H
  - i \text{ sgn } (Q_F) \rho^F_{ij} A \Big]  P_R\; F_{j}\notag \\
 &-\overline{U}_i\left[(V\rho^D)_{ij} P_R-(\rho^{U\dagger}V)_{ij} P_L\right]D_j H^+ 
- \bar{\nu}_i\rho^L_{ij} P_R \; L_j H^+ + \text{h.c.} \, ,
\label{eq:eff}
\end{align}
where $i,j = 1, 2, 3$ are generation indices, $P_{L,R}\equiv
(1\mp\gamma_5)/2$, and $V$ is the CKM matrix. In flavor space, the
fermion fields $F$ are defined as $U=(u,c,t)$, $D = (d,s,b)$,
$L=(e,\mu,\tau)$ and $\nu=(\nu_e,\nu_\mu,\nu_\tau)$. While the mass
matrices are diagonalized as in the Standard Model, one cannot rotate
away $CP$-violating phases of the second set of $\rho^F$ matrices in
the general 2HDM even in the Higgs basis.  That is, the two coupling
matrices are
\begin{align}
  \lambda^F_{ij} =\sqrt{2} \, \frac{m_i^F}{v} \delta_{ij} \in \mathbb{R}
  \qquad \text{and} \qquad 
  \rho^F_{ij} \in \mathbb{C} \; .
\end{align}
The complex coupling matrices $\rho^F$ are, strictly speaking, not
related to the fermion masses. On the other hand, given experimental
constraints and a possible order-of-magnitude correspondence in the
values of $\rho^F$ and $\lambda^F$ lead us to consider $\rho^U_{tj}$
or $\rho^U_{tt}$.

\begin{figure}[t]
  \center
  \includegraphics[width=.495\textwidth]{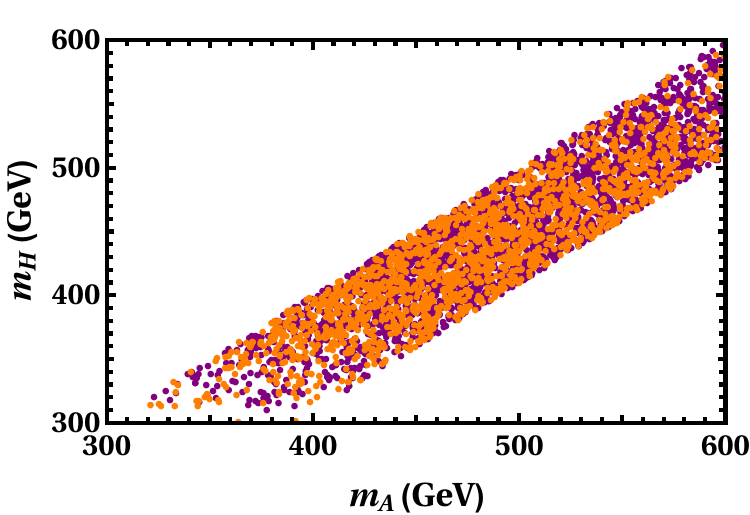}
  \includegraphics[width=.495\textwidth]{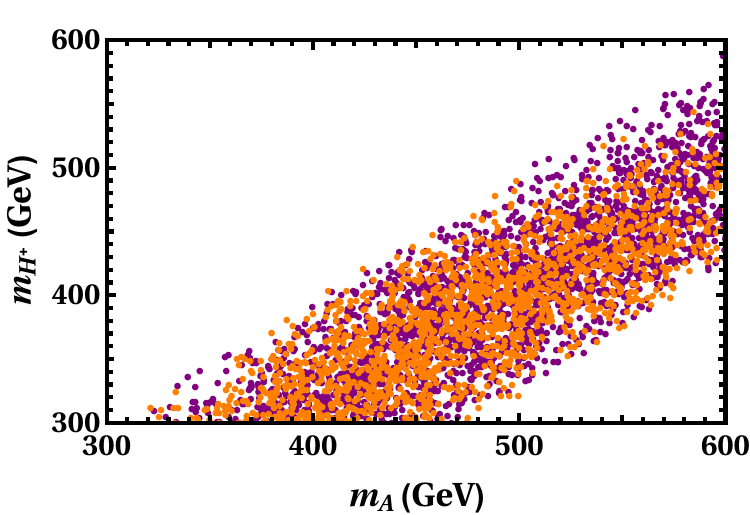}
  \caption{Parameter space allowed by perturbativity, positivity,
    unitarity, and electroweak precision measurements in the
    $m_A$--$m_H$ and $m_A$-$m_{H^\pm}$ planes. The purple and orange
    scanned points corresponds to $c_\gamma =0.1$ and 0.3
    respectively.}
  \label{fig:mass}
\end{figure}

In principle, a complex $\rho_{tt}$ can robustly drive
EWBG~\cite{Fuyuto:2017ewj}, which motivates search for channels like
$gg\to H\to t \bar t$ or $gg\to H t \bar t\to
4t$~\cite{Craig:2015jba,Kanemura:2015nza,Gori:2016zto}.
In this paper we focus instead on complex off-diagonal entries
$\rho_{tj}$, specifically $\rho_{tc}$. With a large phase, this FCNC
coupling can also account for the observed baryon
asymmetry~\cite{Fuyuto:2017ewj}. One of its merits is that $\rho_{tc}$
does not generate an electron EDM through the
Barr-Zee~\cite{Barr:1990vd} two-loop mechanism, and can therefore more
easily~\cite{Fuyuto:2019svr} evade the ACME
bound~\cite{Andreev:2018ayy} $d_e < 1.1 \times 10^{-29}\; e$\,cm.
Moreover, if we assume $\rho_{ct}$ to be small, the constraint on
$\rho_{tc}$ from the charm chromo-EDM also
vanishes~\cite{Sala:2013osa}. We therefore define our specific
baryogenesis scenario as~\cite{Fuyuto:2017ewj}
\begin{align}
  |\rho_{tc}|\gtrsim 0.5
  \qquad \text{and} \qquad
  |c_\gamma| \gtrsim 0.1 \; ,
\label{eq:paras}
\end{align}
with a sufficiently large complex phase.  A strong first-order phase
transition is then possible
for~\cite{Funakubo:1993jg,Cline:1996mga,Kanemura:2004ch,Fromme:2006cm,
  Borah:2012pu,Cline:2013bln,Dorsch:2013wja,Blinov:2015vma,Fuyuto:2015jha,Basler:2016obg}
\begin{align}
  m_{A,H,H^+} \sim 300~...~600~\gev.
\end{align}
This mass range is allowed by perturbativity, positivity, unitarity,
and electroweak precision data. We rely on
2HDMC~\cite{Eriksson:2009ws} to provide the results of
Fig.~\ref{fig:mass} for $c_\gamma = 0.1$ and 0.3.  The 2HDMC
parameters in the Higgs basis are $\eta_{1,..,7}$ and $m_{H^\pm}$.  To
save computing time we actually scan 
$\mu_{22} \in [0, 1000]$~GeV,
$m_A \in [300, 600]$~GeV,
$m_H \in [300, 600]$~GeV,
$m_{H^\pm} \in [300, 600]$~GeV, 
$\eta_2 \in [0, 6]$, and $\eta_7 \in [-6, 6]$, and 
express them in terms of the $\eta_i$.
To match the 2HDMC conventions we define $\gamma\in
[-\pi/2,\pi/2]$. We refer readers to
Refs.~\cite{Hou:2019qqi,Hou:2019mve,Modak:2019nzl,Modak:2020uyq,Ghosh:2019exx}
for further details on the parameter scan.


\begin{figure}[t]
  \center
  \includegraphics[width=.495 \textwidth]{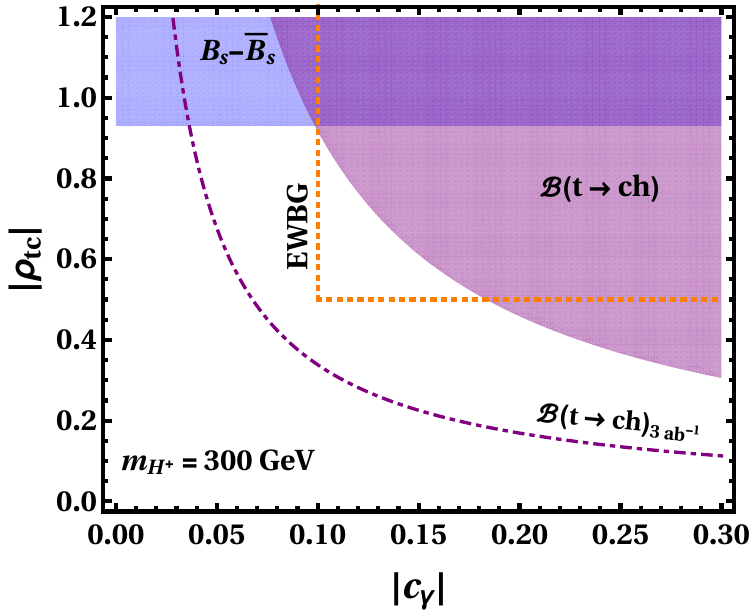}
  \includegraphics[width=.495 \textwidth]{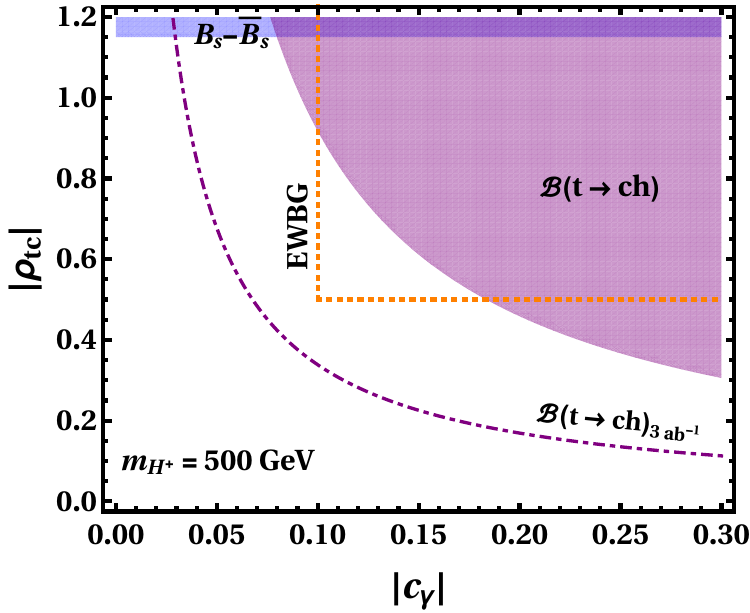}
  \caption{Indirect constraints from $B_s-\overline{B}_s$ mixing (blue), 
    $\br(t\to c h)$ (purple) in the 
    $\rho_{tc}$--$c_\gamma$ plane for two $H^+$ mass values, together with the 
    baryogenesis region (orange).}
  \label{fig:low}
\end{figure}

As the first constraint on $c_\gamma$ and the set of $\rho_{ij}$ we
consider measurements of the SM-like Higgs.  Higgs coupling
measurements constrain the Higgs mixing angle to $c_\gamma \leq 0.3$
and 95\%CL.  Our choice of $\rho_{tc}$ as the source of $CP$-violation
is motivated by its much weaker constraints, because it hardly affects
SM-like Higgs production and decay. The relevant constraints on
$\rho_{tc}$ are indirect.  For flavor observables, $\rho_{tc}$ enters
through loops with charm quarks and a charged Higgs into $B_{s} -
\overline{B}_{s}$ mixing and $\br(B\to X_s\gamma)$. Reinterpreting the
limit from Ref.~\cite{Crivellin:2013wna} we find
\begin{align}
  |\rho_{tc}| \lesssim 1,
  \qquad \text{for} \quad m_{H^+}=300~\gev,
\end{align}
and its counterpart for $m_{H^+}=500$~GeV is illustrated in
Fig.~\ref{fig:low}, alongside with the EWBG-region. The limit is
relatively weak in our general model, in contrast to the type-II 2HDM,
and for larger $m_{H^+}$ it rapidly becomes irrelevant.  Finally,
finite $c_\gamma$ in combination with $\rho_{tc}$~\cite{Chen:2013qta}
leads to anomalous top decays $t \to ch$~\cite{Hou:1991un}, forbidden
at tree level in the SM. The current Run~2 limits at 95\%CL are
\begin{align}
  \br(t\to c h) \approx \frac{c_\gamma^2 |\rho_{tc}|^2}{7.66 + c_\gamma^2 |\rho_{tc}|^2} <
  \begin{cases}
   1.1\times 10^{-3}
  \qquad \text{ATLAS~\cite{Aaboud:2018oqm}} \\
   4.7 \times 10^{-3}
  \qquad \text{CMS~\cite{Sirunyan:2017uae}} \, .
  \end{cases}
\end{align}
They get weaker for smaller $c_\gamma$ and vanish in the alignment
limit.  We illustrate the stronger ATLAS~\cite{Aaboud:2018oqm}
constraint also in Fig.~\ref{fig:low}, along with the projected HL-LHC
95\%CL upper limit $\br(t\to c h)< 1.0 \times
10^{-4}$~\cite{ATL-PHYS-PUB-2016-019}. While an observation of this
anomalous decay could point to a large value of $|\rho_{tc}|$, if
would not provide a link to baryogenesis. A natural step towards
solving the baryogenesis puzzle would be to search for new scalar
degrees of freedom related to this flavor-changing coupling.

While we will focus on $\rho_{tc}$ throughout this paper, we point out
that $\rho_{tu}$ can be tested using a very similar strategy. For the
LHC processes discussed in the coming sections, there is always a
corresponding process with an up-quark replacing the charm-quark.  One
difference between the two FCNC scenarios is that $\rho_{tu}$ can
induce observable effects in $\br(B\to \mu \nu)$~\cite{Hou:2019uxa},
within the reach of Belle-II~\cite{Kou:2018nap}. The combination of
$\rho_{tc}$ and $\rho_{tu}$ is subject to very strong constraints from
$D$--$\overline{D}$ mixing~\cite{Crivellin:2013wna}, and we will
assume only one of the two, but not both at the same time.

\section{Charged Higgs production}
\label{sec:bwh}

\begin{figure}[t]
  \center
  \includegraphics[width=.6\textwidth]{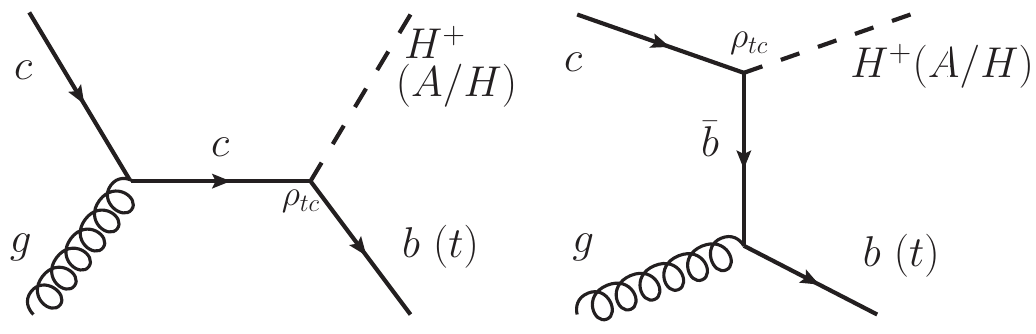}
  \caption{Leading-order Feynman diagrams for the $\rho_{tc}$-induced
    $cg \to b H^+$ and $cg\to t A/tH$ processes.}
  \label{fig:feyn}
\end{figure}

In the EWBG parameter region of Eq.\eqref{eq:paras}, the partonic
process at LHC
\begin{align}
  cg \to b H^+ \to b \, (W_\ell^+ h)
  \to b \; W^+_\ell W^+_\ell W^-_\ell ,
\label{eq:proc_charged}
\end{align}
probes $\rho_{tc}$ in $H^+$-production and $c_\gamma$ in the decay
$H^+ \to W^+ h$. The production benefits from the relatively large
charm density in the proton, as well as the
combination~\cite{Ghosh:2019exx} with the CKM matrix element $V_{tb}$
following Eq.\eqref{eq:eff}. The leading-order Feynman diagrams are
presented in Fig.~\ref{fig:feyn}.  While we will require a tagged
$b$-jet, the $b$-inclusive production process could also be defined as
$c\bar b \to H^+$~\cite{Plehn:2002vy,Boos:2003yi}. For a clean
analysis, we assume that all three $W$-bosons decay to either
electrons or muons.  The same process is induced by $\rho_{ct}$, but
this coupling is constrained to be much
smaller~\cite{Altunkaynak:2015twa} by flavor constraints.

\begin{table}[b!]
\centering
\begin{tabular}{cc|cc|c}
\toprule
$m_{H^+}$[GeV] & \;$\Gamma_{H^+}$[GeV]\; & \,$\br(H^+\to c\bar b)$\, & \,$\br(H^+ \to W^+ h)$\; & \;$\sigma (cg \to bH^+)$\,[fb] \\
\midrule
   350  &  2.2  & 0.85 & 0.15 & 0.126 \\
   500  &  3.9  & 0.66 & 0.34 & 0.113 \\
\bottomrule
\end{tabular}
\caption{Charged Higgs properties for the two benchmark points with
  $\rho_{tc}=0.35$ and $c_\gamma=0.25$. The quoted LHC cross sections
  include the decay $H^+ \to Wh$ in the fully leptonic mode, as shown
  in Eq.\eqref{eq:proc_charged}, as well as selection and background
  rejection cuts.}
\label{tab:sig}
\end{table}

The $H^+ W^- h$ coupling, modulated by $c_\gamma$, arises
from~\cite{Djouadi:2005gi,Branco:2011iw}
\begin{align}
\lag \supset -\frac{g_2}{2} c_\gamma \left(h \partial^\mu H^+
         -  H^+ \partial^\mu h   \right) W^-_\mu + \text{h.c.} \, , 
\end{align}
where $g_2$ is the $SU(2)$ gauge coupling.  To estimate the reach of
our charged Higgs signal, we choose two allowed benchmark points,
\begin{align}
  \rho_{tc}=0.35, \qqquad c_\gamma=0.25, \qqquad 
  m_{H^+} = 350, 500~\gev,
\end{align}
as given in Tab.~\ref{tab:sig}.  For the branching ratios, we ignore
the loop-induced decays $H^+ \to W^+ \gamma$ and $H^+ \to W^+ Z$.  We
generate signal and background events for $\sqrt{s}=14$~TeV at leading
order with MadGraph5\_aMC@NLO~\cite{Alwall:2014hca}. The effective
model is implemented in the FeynRules~\cite{Alloul:2013bka} framework,
and for parton densities we use NN23LO1~\cite{Ball:2013hta}. The
events are showered and hadronized with
PYTHIA6.4~\cite{Sjostrand:2006za} and then handed to
Delphes~3.4.2~\cite{deFavereau:2013fsa} for a fast detector simulation
with the default ATLAS card. Jets are reconstructed with an $R=0.6$
anti-$k_T$ algorithm~\cite{Cacciari:2008gp} in
FastJet~\cite{Cacciari:2011ma}. For $b$-tagging as well as $c$-jet and
light-jet rejections, we also rely on the default ATLAS card. To allow
for extra jets we apply MLM
matching~\cite{Mangano:2006rw,Alwall:2007fs} with the default
MadGraph5\_aMC run card. The signal is generated with up to two
additional jets, do account for higher-order effects in the event
kinematics.

\begin{table}[t]
\begin{small}
\begin{tabular}{c|ccccccccc|c}
\toprule
& $t\bar tW$  & \ $t t\bar Z$ & $WZ\,+\,$jets & $4t$ & \ $t\bar t h$ & $t Z$\,+\,jets & \,$tWZ$ & $ZZ+$jets\, & $t\bar t+$jets & \,sum bkg  \\
\midrule                                
merged jets\,              & 1     & \ 1     & 1     & 0     & \ 0     & 1     & \,0    & 1     & 1 & \\ 
$K$-factor               & NLO   & \ NLO   & NNLO  & NLO   & \ NLO   & NLO   & \,LO   & LO & NNLO & \\
$\sigma_\text{bkg}$~[fb] & \, 0.685 & \ 0.279 & 0.101 & 0.074 & \ 0.026 & 0.017 & \,0.02 & 0.001 & 0.304 & 1.504 \\
\bottomrule
\end{tabular}
\end{small}
\caption{Background cross sections for the charged Higgs process after
  cuts.}
\label{tab:bkg}
\end{table}

The dominant SM-backgrounds are $t\bar t W$ and $t\bar t Z$
production, followed by $WZ$\,+\,jets, $4t$, $t\bar t h$, $tZj$,
$tWZ$, and $ZZ\,+\,$jets.  Furthermore, we find the backgrounds $3t$,
$3t+W$, and $3W$ to be negligible, so we ignore them in our analysis.
However, given a mis-identification probability for a jet as a lepton
around $10^{-4}$~\cite{Aad:2016tuk,Alvarez:2016nrz}, $t\bar t$
production will lead to non-trivial background contributions.  For all
backgrounds, we use the same simulation chain as for the signal, with
up to one additional jet for $t\bar t W$, $t\bar t Z$, $WZ\,+\,$jets,
$ZZ$\,+\,jets, $tZ$\,+\,jets, $t\bar t$+\,jets, and no QCD jets for
the high-multiplicity backgrounds $4t$, $tWZ$ and $t\bar t h$.  To
approximately account for QCD corrections in addition to the jet
emission, we attach NLO $K$-factors to the dominant $t\bar t V$
backgrounds, namely 1.35 ($W^-$), 1.27 ($W^+$)~\cite{Campbell:2012dh},
and 1.56 ($Z$)~\cite{Campbell:2013yla}. We also correct the
$WZ\,+\,$jets and $t\bar t$+\,jets background normalizations to NNLO
by factors 2.07~\cite{Grazzini:2016swo} and 1.84~\cite{twikittbar}
respectively. Furthermore, we adjust the $4t$, $t\bar t h$, and $\bar
tZ\, +\,$jets rates to NLO through the $K$-factors
2.04~\cite{Alwall:2014hca}, 1.27~\cite{twikittbarh} and
1.44~\cite{Alwall:2014hca}.  The cross sections for the signal and
$tWZ$ are kept at LO for simplicity.  Here, we simply assume the QCD
correction factors for the $W^+Z\,+\,$jets and $tZ\,+\,$jets processes
to be the same as their respective charge-conjugate processes.

To suppress the backgrounds, we adopt a simple set of requirements.  We
start with events containing at least three charged leptons and at
least one tagged $b$-jet passing
\begin{alignat}{7}
  p_{T,\ell} &> 20~\gev,   & \qqquad   |\eta_\ell| &< 2.5, \notag \\
  p_{T,b} &> 20~\gev,     & \qqquad   |\eta_b| &< 2.5, \notag \\
  \Delta R_{ij} &> 0.4, \qquad (i,j=\ell,b) \notag \\
  \met &> 35~\gev,  & \qqquad 
  m_{\ell^+ \ell^-} &\not \subset [ 76, 110]~\gev, \qquad (\ell = e,\mu)  \, .
\end{alignat}
The same-flavor opposite-sign dilepton veto reduces the dominant
$t\bar t Z$ background.  In case more than one such $\ell^+ \ell^-$
pair exists, we select the combination closest to the $Z$-mass for
rejection. The remaining signal rate is given in Tab.~\ref{tab:sig},
while the background rates are summarized in Tab.~\ref{tab:bkg}.

For discovery reach and exclusion limits, we compute the significance using
the likelihood for a simple counting experiment~\cite{Cowan:2010js}.
If we observe $n$ events with  $n_\text{pred}$ predicted, the agreement
between observation and prediction is given by
\begin{align}
  Z(n|n_\text{pred})= \sqrt{-2\ln\frac{L(n|n_\text{pred})}{L(n|n)}},
  \qquad \text{with} \qquad L(n|\bar{n}) = \frac{e^{-\bar{n}} \bar{n}^n}{n !} \, \label{poisso}.
\end{align}
For discovery, we compare the observed signal plus background with the
background prediction and require $Z(s+b|b) > 5$. For exclusion, we
assume a background-consistent measurement after predicting a signal
on top of the background, such that $Z(b|s+b) > 2$. For instance,
assuming an HL-LHC data set with $3000~\ifb$ and the signal and
background cross sections in Tabs.~\ref{tab:sig} and \ref{tab:bkg}, we
find a significance of $\sim5.6\sigma$ for $m_{H^+}=350$~GeV and
$\sim5\sigma$ for $m_{H^+} = 500$~GeV.

\begin{figure}[t]
  \center
  \includegraphics[width=.495\textwidth]{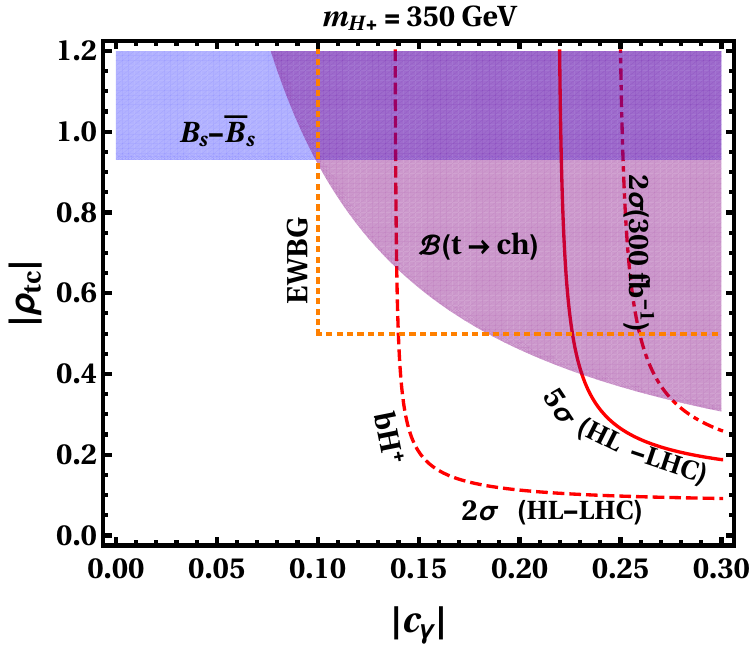}
  \includegraphics[width=.495\textwidth]{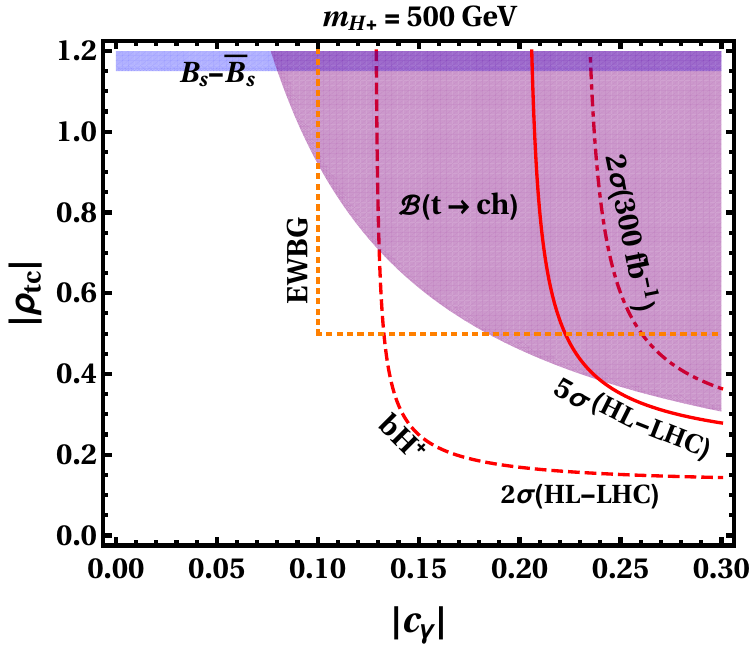}
  \caption{Projected $300~\ifb$ exclusion (dot-dashed) and HL-LHC 
  discovery (solid) and exclusion (dashed) contours for the charged Higgs
    signature $pp \to bH^+ \to b W^+ h$, along with EWBG-favored region
    and the indirect constraints from Fig.~\ref{fig:low}.}
  \label{fig:charged}
\end{figure}

We illustrate in Fig.~\ref{fig:charged} the Run~3 and HL-LHC reach for
the charged Higgs signature in the $|\rho_{tc}|$--$c_\gamma$ plane.
We see from the left panel that Run~3 can exclude $|\rho_{tc}| > 0.3$
and $|c_\gamma| = 0.27$ for $m_{H^\pm}=350$~GeV, while the HL-LHC will
be sensitive to $|\rho_{tc}| > 0.2$ and $|c_\gamma| = 0.14$. For
larger Higgs masses, the expected limits become only slightly weaker.
The $b$-associated charged Higgs channel covers the $|\rho_{tc}|$
range preferred by EWBG, but there remains a slice of EWBG parameter
space with $|c_\gamma| \lesssim 0.14$.  This follows as an effect of
decreasing $\br(H^\pm \to W^\pm h)$ with smaller $c_\gamma$.
Unfortunately, this hole is unlikely to be filled by other charged
Higgs decays, because for instance the standard signature $H^+ \to
t\bar{b}$ requires large production rates.  Here, utilizing the
expression from Ref.~\cite{Crivellin:2013wna}, the limit from $B_s$ in
the left panel of Fig.~\ref{fig:charged} is plotted for $m_{H^+}= 350$
GeV to conform with the benchmark charged Higgs mass for $pp \to bH^+
\to b W^+ h$ signature.

\section{Neutral Higgs production}
\label{sec:sstop}

To cover the parameter region $|c_\gamma| < 0.14$, left open by the
charged Higgs signature, we turn to the neutral Higgs channel,
\begin{align}
  cg \to t H/t A \to t \; (t \bar{c}) \, ,
\end{align}
also given in Fig.~\ref{fig:feyn}, where production and decay are
both mediated by $\rho_{tc}$. A very slight $c_\gamma$-dependence of
the $cg\to tH/tA \to t t \bar c$ process arises from the heavy Higgs
branching ratios. Non-resonant and $t$-channel diagrams with $H/A$
exchange leading to $cc\to tt$ scattering as well as $gg\to t t \bar c
\bar c$, though small, are included in our signal analysis.

For small $c_\gamma$, the neutral Higgs production process currently
leads to the most stringent limit on
$\rho_{tc}$~\cite{Hou:2018zmg,Hou:2019gpn}, because it affects the SM
control region of the Run~2 $tt\bar t\bar t$ ($4t$) analysis by
CMS~\cite{Sirunyan:2019wxt}.  Based on the number of $b$-jets and
leptons, CMS divides its analysis into several signal and two control
regions.  The most stringent constraint on $\rho_{tc}$ arises from the
$t\bar t W$ control region (CRW)~\cite{Kohda:2017fkn,Hou:2018zmg}. The
CMS baseline selection includes two same-sign leptons with
\begin{align}
  p_{T,\ell} > 25,20~\gev
  \qquad \text{and}\qquad
  |\eta_e| < 2.5, \qquad |\eta_\mu| < 2.4 \, ,
\end{align}
where the charge-misidentified Drell-Yan background is reduced by
vetoing same-sign electron pairs with $m_{ee} <12$~GeV. The CRW then
requires two to five jets, two of them $b$-tagged.  All jets have to
fulfill $|\eta_j| < 2.4$, and events are selected if they fulfill any
one of
\begin{alignat}{7}
  \text{(i)} &\qquad &
  p_{T,b_1} &> 40~\gev, \qquad  p_{T,b_2} > 40~\gev, \notag \\
  \text{(ii)} &\qquad &
  p_{T,b_1} &> 20~\gev, \qquad p_{T,b_2} = 20~...~40~\gev, \qquad & p_{T,j_3} &> 40~\gev, \notag \\
  \text{(iii)} &\qquad &
  p_{T,b_{1,2}} &= 20~...~40~\gev, \qquad & p_{T,j_{3,4}} &> 40~\gev.
\label{eq:neutral_123}
\end{alignat}
Finally, the analysis requires~\cite{Sirunyan:2019wxt}
\begin{align}
  H_T = \sum_\text{jets} p_{T,j} > 300~\gev
  \qquad \text{and} \qquad
  \met > 50~\gev.
\end{align}
With this selection, CMS observes 338 events with $335 \pm 18$ events
expected from SM-backgrounds plus $4t$ signal.  To estimate the CRW
limits on $\rho_{tc}$, we generate both neutral Higgs processes with
the decay $H/A \to t \bar c$, followed by lepton-hadron combinations
of the top decays at $\sqrt{s}= 13$~TeV. We use the same setup as for
the charged Higgs simulations, except that we use the default CMS
detector card in Delphes~3.4.2. Remaining uncertainties on our
simulation affect the $c$-initiated processes $cg \to bH^+$ and $cg
\to t A/t H$, such as from parton densities and scale
dependence~\cite{Buza:1996wv,Boos:2003yi,Maltoni:2012pa,Butterworth:2015oua}. We
expect them to be small, and do not include them, just as we do not
account for non-prompt and fake backgrounds.

There exist a similar ATLAS search~\cite{Aad:2020klt}, but it is less
constraining~\cite{Hou:2020ciy}.  This is primarily due to the
definition of signal regions and selection criteria. Furthermore,
searches for squark pair production in $R$-parity violating
supersymmetry~\cite{Aad:2019ftg} and exotics searches for same-sign
dileptons and $b$-jets~\cite{Aaboud:2018xpj} involve similar final
states, but their selection cuts are too model-specific to be applied
to our signature.

To judge the impact of the existing CMS CRW limits from $4t$ search,
we focus on the border of the EWBG-region with $c_\gamma = 0.1$ and
$|\rho_{tc}|= 0.5$. We stick to our two charged Higgs masses, assume
$m_A \approx m_{H^\pm} = 350, 500$~GeV for the pseudoscalar, and
decouple the heavy scalar $H$.  In this scenario, the same-sign top
contribution to the CRW arises from $cg\to tA\to t t \bar c$.  We
demand that the combination of SM-backgrounds and heavy neutral Higgs
production agree with observed within $2\sigma$ and give the excluded
regions in Fig.~\ref{fig:neutral}. To scan the parameter space we use
a simplified scaling $|\rho_{tc}|^2\br (A\to t \bar c)$, such that
$\Gamma_A = 3.05\, (6.08)$~GeV for $m_A= 350\, (500)$~GeV.  The
exclusion covers most of the EWBG-region except for small values of
$|\rho_{tc}|$. 

A dedicated same-sign top search, such as the $pp\to tA + X \to t
t\bar c + X$ study of Ref.~\cite{Hou:2020ciy}, can probe the nominal
parameter space of $\rho_{tc}$-EWBG.  This process can be searched
for in events containing same-sign dileptons ($ee$, $\mu\mu$, $e\mu$),
at least three jets with at least two $b$-tag, and some $\met$. The
dominant backgrounds are $t\bar t Z$, $t\bar t W$, $4t$, while $t\bar
t h$, with $tZ\, +$\,jets, $3t + W$ and $3t + j$ give subdominant
contributions, and the non-prompt background can be 1.5 times the rate
of $t\bar t W$. In addition, if a lepton charge gets misidentified,
the $t\bar t\,+$\,jets and $Z/\gamma^*\,+$\,jets processes will also
contribute. For further details of the QCD correction factors for
different backgrounds, we refer to Ref.~\cite{Hou:2020ciy}.  To reduce
backgrounds, we applied an event selection different from the CRW of
Ref.~\cite{Sirunyan:2019wxt}: the leading and sub-leading same-sign
leptons should have $p_T > 25 (20)$~GeV and $|\eta| < 2.5$.  All three
jets are required to have $p_T> 20$~GeV and $|\eta| < 2.5$. All jets
and leptons are separated by $\Delta R_{ij} > 0.4$.  The all event
should have $\met > 35$~GeV.  and $H_T > 300$~GeV, where the latter
includes the two leading sames-sign leptons. The background cross
sections after selection cuts are summarized in
Tab.~\ref{tab:backgsst}.

\begin{table}[t]
\centering
\begin{tabular}{cc|cc|cc}
\toprule
\,background                &  ~$\sigma$\,[fb] & \,background                &  ~$\sigma$\,[fb] & \,background                &  ~$\sigma$\,[fb] \\
\midrule 
                       $t\bar{t}W$                & 1.31~  &
                       $t\bar{t}Z$                &  1.97~    &
                       ~$tZ\,+$\,jets                & \,\ 0.007                \\
                       $4t$                       & \,\ 0.092~  &
                       $3t+W$                     & \,\ 0.001~   &
                       ~$3t+j$                     & \,\ \ \,0.0004      \\
                       $t\bar t h$                & \,\ 0.058~     &
                       charge-flip                   & \,\ 0.024~        &
                       ~non-prompt                  & ~~~$1.5\times$ $t\bar{t}W$ \\
\bottomrule
\end{tabular}
\caption{Background cross sections for the dedicated
  same-sign top search after selection cuts at $\sqrt{s}=14$\;TeV.}
 \label{tab:backgsst}
\end{table}

For the reference values $|\rho_{tc}|=0.5$ and $c_\gamma = 0.1$, we
generate the same-sign top cross sections for $m_A=350$ and $500$ GeV.
Based on the background rates of Tab.~\ref{tab:backgsst} and
Eq.\eqref{poisso}, rescaling the signal cross section by
$|\rho_{tc}|^2\br(A\to t \bar c)$, we find the exclusion (green
dashed) and discovery (green solid) contours in the
$|c_\gamma|$--$|\rho_{tc}|$ plane as given in Fig.~\ref{fig:neutral}.

\begin{figure}[t]
  \center
  \includegraphics[width=.495\textwidth]{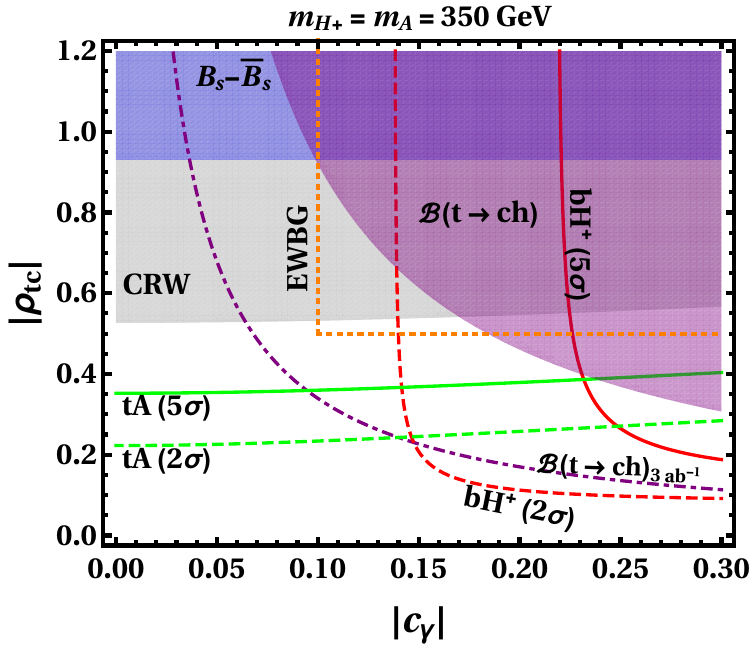}
  \includegraphics[width=.495\textwidth]{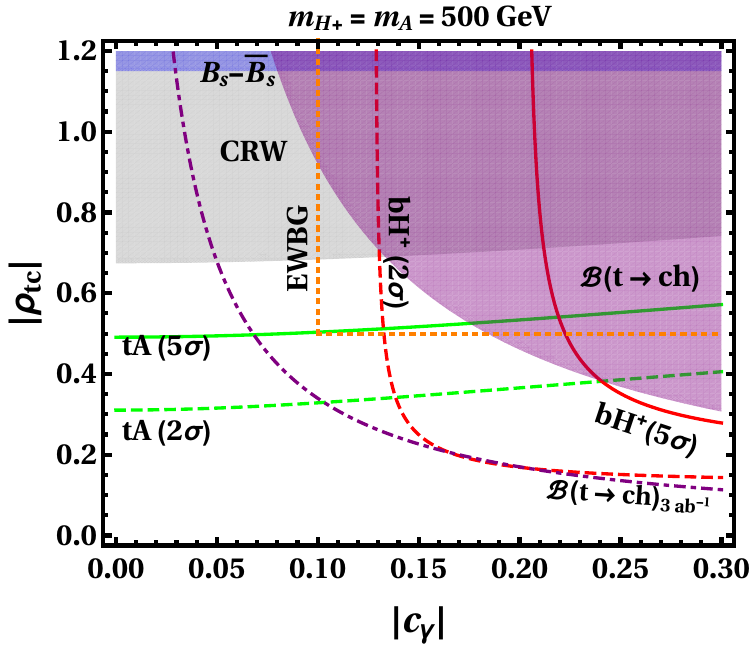}
  \caption{Exclusion regions from neutral Higgs production
    contributing to the CMS CRW~\cite{Sirunyan:2019wxt} (gray shades), as
    well as HL-LHC expectations from a dedicated same-sign top
    search~\cite{Hou:2020ciy} (green). We also show the EWBG region
    and the indirect constraints from Fig.~\ref{fig:low} and the
    HL-LHC charged Higgs reach from Fig.~\ref{fig:charged}.}
\label{fig:neutral}
\end{figure}

A loop hole in the neutral Higgs analysis appears though the
destructive interference of $cg\to t H \to t t \bar c$ and $cg\to tA
\to t t \bar c$.  If the widths and masses of the two heavy neutral
Higgses become degenerate, the two production processes completely
cancel~\cite{Kohda:2017fkn,Hou:2018zmg} and the same-sign top
signature vanishes. Our limits derived from $A$-production would be
similar for $H$-production with $m_A \gg m_H$.  We now illustrate
limits for $m_A \sim m_H$ with a case where the three heavy Higgs
masses are of similar size, specifically $m_{H^\pm}=350\,(500)$~GeV,
$m_A=343\,(524)$~GeV, and $m_H=355\,(501)$~GeV. The self-couplings are
$\eta_1=0.276\,(0.297)$, $\eta_2 =1.335\,(2.762)$, $\eta_3=
1.66\,(1.21)$, $\eta_4= -0.04\,(0.398)$, $\eta_5 =0.121\,(-0.428)$,
$\eta_6 = -0.181\,(-0.386)$, $\eta_7= 0.605\,(-0.095)$, and
$\mu_{22}^2/v^2= 1.189\,(3.516)$, in agreement with perturbativity,
positivity, unitarity, and electroweak precision
data~\cite{Eriksson:2009ws}.  The relevant decays are $ A \to t \bar
c, Zh$ and $H \to t \bar c, h h, ZZ, WW$, with mild contributions from
the $\lambda_f c_\gamma$-dependent fermionic decays to $b\bar b$ and
$t\bar t$. For $\rho_{tc}= 0.5$ and $c_\gamma =0.1$, the total widths
are $\Gamma_A = 3.28\,(7.37)$~GeV and $\Gamma_H = 2.91\,(6.56)$~GeV,
and the combined contributions to the CRW rates are 0.467~fb and
0.261~fb, corresponding to 64 and 35.8 events. Demanding that the
combination of events expected in the SM and from the neutral Higgs
channels agree within $2\sigma$ of the observed number, we find that
$|\rho_{tc}| = 0.5$ is already excluded for $m_{H^\pm}=350$~GeV and
$c_\gamma =0.1$, and barely allowed for $m_{H^\pm}=500$~GeV.  We see
that, due to the choice of parameters, the cancellation between $cg\to
t H \to t t \bar c$ and $cg\to tA \to t t \bar c$ is not exact, and
the CRW limit is stronger than the $H$ (or $A$) decoupled case.

As mentioned in the introduction, we ignore all $\rho_{ij}$ couplings
except for $\rho_{tc}$, so before closing we should discuss the impact
of this assumption.  It may well be that the $\rho^{U,D,L}$ matrices
share the flavor-ordering of the Yukawa couplings, $\rho_{tt}\sim
\lambda_t$, $\rho_{bb}\sim \lambda_b$ and $\rho_{\tau\tau}\sim
\lambda_\tau$.  Current data still allows $\rho_{tt} \lesssim
0.5$~\cite{Ghosh:2019exx} and $\rho_{bb}\sim
0.1$~\cite{Modak:2019nzl,Modak:2020uyq} for sub-TeV scalars, and both
parameters can account for the observed baryon asymmetry. The extra
top Yukawa coupling $\rho_{tt}$ can be searched for in signatures such
as $gg \to A/H \to t \bar t$~\cite{Aaboud:2017hnm,Sirunyan:2019wph}
$gg \to A/H t \bar t \to t \bar t t \bar t$~\cite{Sirunyan:2019wxt}
and $g b \to \bar t H^+ \to \bar t t \bar
b$~\cite{Plehn:2002vy,Boos:2003yi,ATLAS-CONF-2020-039,Sirunyan:2020hwv},
while rare decays $\br(B\to X_s \gamma)$ and $B_{d,s}$ mixing provide
indirect probes~\cite{Altunkaynak:2015twa}. In general, a large value
for $\rho_{tt}$ dilutes the decays $A/H \to t \bar c$ and $H^+ \to W^+
h$ through $A/H\to t \bar t$ and $H^+\to t \bar b$.  However, the
combination with $\rho_{tc}$ opens additional discovery modes such as
$cg\to tA/tH\to t t \bar t$~\cite{Kohda:2017fkn} and $cg\to b H^+ \to
b t \bar b$~\cite{Ghosh:2019exx}.  There also exist several direct and
indirect constraints on
$\rho_{bb}$~\cite{Modak:2019nzl,Modak:2020uyq}. Finally, a large
allowed value of $\rho_{tu}$~\cite{Hou:2020ciy} combined with
non-vanishing $\rho_{tc}$ will be constrained by $D$-meson
mixing~\cite{Crivellin:2013wna,Altunkaynak:2015twa}. Similarly,
constraints on $\rho_{tt}$, $\rho_{bb}$, $\rho_{\tau \tau}$ from
flavor physics and low energy observables as discussed in
Refs.~\cite{Crivellin:2013wna,Altunkaynak:2015twa,Iguro:2017ysu,Modak:2020uyq},
and their detailed impact on the $\rho_{tc}$-EWBG would be an
interesting future direction.

\section{Outlook}
\label{sec:disc}

Electroweak baryogenesis is an attractive target for experimental
analysis, because it can be tested by a variety of
measurements. Specific models typically combine new bosonic degrees of
freedom with extra $CP$-violation. In our case, the new degrees of
freedom are provided by a general or type-III 2HDM. If the Higgs
self-couplings are sufficiently large, the heavy Higgs states can be
relatively heavy, so we use $m_{H^+} = 350$ and 500~GeV as benchmark
scenarios. The complex phase is given by an FCNC top--charm coupling
with $|\rho_{tc}| \gtrsim 0.5$, combined with a $CP$-even Higgs mixing
angle $c_\gamma \gtrsim 0.1$. At the LHC, $\rho_{tc}$ has the
advantage that we can test it in processes mediated by this large top
Yukawa, but with a charm quark in the initial state, while it easily
evades EDM constraints.

In the allowed 2HDM parameter space, the charged Higgs has to be
relatively light, which means we can search for it via $cg \to b H^+$
with a subsequent $H^+ \to W^+ h$ decay. Our proposed analysis is
relatively straightforward and probes most of the EWBG parameter space
at the HL-LHC, with the exception of small values of $c_\gamma \sim
0.1~...~0.12$, when $H^+ \to W^+ h$ decay becomes too suppressed by
$CP$-even Higgs boson mixing.

A complementary channel that can survive small $CP$-even Higgs boson
mixing is heavy neutral Higgs production, $cg\to tA/ tH$, together
with $A/H \to t \bar{c}$ decay. In this case, production and decay are
both mediated by $\rho_{tc}$ without being suppressed by small
$c_\gamma$, providing strong limits on $\rho_{tc}$ even for small
$c_\gamma$ values.The search channel at the LHC is same-sign top
pairs, allowing us to extract limits already from Run~2.  At the
HL-LHC, the decay $t \to ch$, charged heavy Higgs searches, and
neutral heavy Higgs searches guarantee a comprehensive coverage of the
$\rho_{tc}$-EWBG parameter space in the general 2HDM, leaving us with
the challenge of observing the $CP$-violating phase in a dedicated
analysis.

\begin{center} \textbf{Acknowledgments} \end{center}

First, we would like to thank Kai-Feng (Jack)~Chen for discussions and
for clarifications on Eq.\eqref{eq:neutral_123}.  We are also grateful
to Eibun~Senaha and Margarete M\"uhlleitner for very helpful
discussions and comments.  WSH is supported by MOST
109-2112-M-002-015-MY3 of Taiwan and NTU 109L104019.  TM is supported
by a Postdoctoral Research Fellowship from Alexander von Humboldt
Foundation.  The research of TP is supported by the Deutsche
Forschungsgemeinschaft (German Research Foundation) under grant
396021762 -- TRR~257 \textsl{Particle Physics Phenomenology after the
  Higgs Discovery}.
  
\bibliography{literature}

\end{document}